\documentclass[%
 reprint,
 amsmath,amssymb,
 aps,
floatfix,
]{revtex4-1}

\usepackage{graphicx}
\usepackage{dcolumn}
\usepackage{bm}

\usepackage{subcaption}
\begin{document}

\preprint{arxiv.org/abs/2001.00034}

\title{Contributions of Talent, Perspective, Context and Luck to Success\\}

\author{Bernardo Alves Furtado}
 \altaffiliation[]{National Council of Research -- CNPq}
 \email{Bernardo.Furtado@ipea.gov.br}
\affiliation{%
 Institute for Applied Economic Research -- Ipea \\
 Brasília, Brazil \\
}%
\date{\today}

\begin{abstract}
We propose a controlled simulation within a competitive sum-zero environment as a \textit{proxy} for disaggregating components of success. Given a simulation of the Risk board game, we consider (a) Talent to be one of three rule-based strategies used by players; (b) Context as the setting of each run of the game with opponents' strategies, goals and luck; and (c) Perspective as the objective of each player. Success is attained when a first player conquers its goal. We simulate 100,000 runs of an agent-based model and analyze the results. The simulation results strongly suggest that luck, talent and context are all relevant to determine success. Perspective -- as the description of the goal that defines success -- is not. As such, we present a quantitative, reproducible environment in which we are able to significantly separate the concepts, reproducing previous results of the literature and adding arguments for context and perspective. Finally, we also find that the simulation offers insights on the relevance of resilience and opportunity.
\end{abstract}

\maketitle

\section{\label{sec:sec1}Introduction}

Western culture has come to believe that success is mostly a consequence of talent and skills, basically derived from meritocratic efforts. A number of papers have shown quite the opposite. In most cases, luck is dominant or the sole effect. Personal characteristics seem to be marginal and restricted to a few cases. We suggest that one of the reasons why the issue is still controversial lies in the difficulty to separate the effects of each one of the components of success. We propose a controlled simulation that enables the distinction among context, talent, perspective, and luck. We do so simulating 100,000 runs of a version of the Risk board game that enables accounting for each element influence towards success.

Risk is a competitive sum-zero game in which players wage war to conquer opponents' countries towards specific goals. Goals are designed so that its completion demands conquering at least eleven new countries or entirely destroying an opponent. Every player starts with seven countries. Players take turns deploying a rule-based number of armies, attacking (and defending) and rearranging troops.

Different environments for each run promotes the environment in which the mixture of talent, luck and perspective contributes to success. We apply the following definitions:

\begin{enumerate}
    \item \textbf{Talent} is a rule-based technique. The ability or skill that each one of the individual strategies within the simulation has to lead to a successful winner. As the results will demonstrate, there is a clear performance distinction among the strategies. Such configuration provides that each different strategy has a clearly defined, permanent, quantifiable talent.
    \item \textbf{Perspective} is defined as the goal that each player has to achieve. Once more the analysis of the simulation will show that there are easier and more difficult objectives in the game. That indicates that a given effort, or luck in a given run may be sufficient to determine a winner under a given goal, but insufficient on another one. 
    \item \textbf{Context} considers that each game run may vary in difficulty as the other five players have ever-changing luck, strategies and goals. Basically, results are not independent \cite{albert_curve_2007}. That means that players with exactly the same strategy and the same amount of luck would perform differently, conditionally on each game sets of other players' strategies, goals and luck. Further, because Risk is a competitive, single-winner game, success only comes when the other players are unable to achieve their goals within a shorter number of runs. Finally, context may also include various kinds of cooperation. One player with a destroy goal may be helped by an opponent. Or two players may independently and without communication or collusion help each other against a third power.  
    \item\textbf{Luck} is a chance occurrence \cite{mauboussin_success_2012}. It derives from rolling dice in the form of a \texttt{numpy python} pseudo-random generator. Luckily, however, given the fact that we are running a simulation, we are able to record each time a die is rolled, and save the results considering the opponent's die. For example, getting a two when the defender gets a one may be just lucky enough. However, having a six when the defender also draws a six is bad luck, as rules determine that ties always favor the defender. 
\end{enumerate}{}

We follow Ioannidis \cite{ioannidis_why_2005} precepts of an evidence-based analysis. We use definitions with a clear design and implementation whilst having no interest in the results. We also are able to run games with long trajectories, incorporating sample size within the discussion. 

Pluchino et al. \cite{pluchino_talent_2018} make a comprehensive review on talent and luck. They apply a simple agent-based model (ABM) to test the assumption that success is a talented-weighted multiplicative result. In their own words "...building on a minimal number of assumptions, i.e., a Gaussian distribution of talent and a multiplicative dynamics for both successes and failures, we present a simple model..." \cite[p.1850014-4]{pluchino_talent_2018} As such, talented people happen to be more successful, but just marginally so, with non-talented people also reaching stardom quite often.

Moreover, Pluchino et al. \cite{pluchino_talent_2018} show that luck and randomness have proven to be decisive to determine success in many areas, such as scientists citation \cite{ruocco_bibliometric_2017}, innovative ideas \cite{iacopini_network_2018}, music \cite{salganik_experimental_2006} and finance \cite{robert_h_frank_success_2016}; or failure, such as a disease interrupting trajectories \cite{tomasetti_stem_2017}. Others \cite{liechti_luck_2014} defend the view that luck is relevant on some occasions, but that even believing in luck, or perception of being luck, might alter ones behavior.

Chamorro-Premuzic \cite{chamorro-premuzic_talent_2016} conducted a survey to explore how much luck is important. He argues that luck is relevant, but talent and effort also play significant roles. That is in line with the findings of \cite{duckworth_grit:_2016}, who conceptualized effort as grit.

This perception that luck may play a lesser role is in accordance with the general view of surveyed CEOs \cite{liechti_luck_2014}. When asked what affects workers performance, talent was found to be more relevant. Thus, a belief in talent and meritocracy remains. Still about perception of luck, Liechti \cite{liechti_luck_2014} argues that the simple belief in luck  changes people's behavior and those who believe tend to be consistently less entrepreneurial. 

This view of unimportance of luck is counterbalanced by Taleb, for instance, who says that "life is the cumulative effect of a handful of significant shocks" \cite[p. xix]{taleb_black_2010} and supported by Pluchino et al. \cite{pluchino_talent_2018}, for whom luck is determinant and talent is just marginal.

In finance, the literature is interested in decomposing talent and luck so that choices and rewards can be better distributed. Deciding on the correct timing and sector to launch new businesses has been said to be a skill \cite{gompers_skill_2006} that can predict future success \cite{gompers_performance_2010}. Whereas when fund performance is concerned, talent is large enough to overcome the costs of paying analysts for just a few number of firms \cite{fama_luck_2010}. Follow-up on Fama and French's paper \cite{korteweg_skill_2017} confirms small influence of managers of Equity Mutual Funds on performance and success.

In the labor market, however, luck is found to be not important, relatively to managerial skill, firm size and market competition \cite{brookman_managerial_2013}. Campbell and Thompson \cite{campbell_why_2015} investigates pay asymmetry for CEOs finding that most difference does not come from luck or skill, but from a number of entangled factors, being executive retention a relevant one. Moreover, Korteweg and Sorensen \cite{korteweg_skill_2017} find that one can only explain firms' performance based on skill alone when having a large sample -- including at least 50 lags -- and details at the level of individual members of the firm in order to predict correct performance.

In terms of different methodologies applied to disaggregating luck and talent, we could mention survey techniques \cite{liechti_luck_2014}, simulation and bootstrapping \cite{yang_luck_2017} and variance decomposition \cite{korteweg_skill_2017}.

Besides the discussion of contributions of skill to performance in finance, a legal question -- whether poker is a game of chance or a game of skill -- has been discussed for some time \cite{cabot_poker_2005,meyer_is_2013,levitt_role_2014}.

Levitt \cite{liechti_luck_2014} states that luck is the "dominant factor", although skill may have marginal influence. Meyer \cite{meyer_is_2013} opted for a controlled experiment to confirm that experts failed to financially outperform average players. 

Simulation of board game Risk has been present in the literature before with emphasis on the learning environment \cite{wolf_intelligent_2005} and as a field in which Artificial Intelligence (AI) players could be tested and enhanced \cite{bence_szabo_risk_2015}. The game has also been used as a means of testing Monte Carlo Tree Search methods \cite{brand_sample_2014} and collaborative systems \cite{badeig_analyzing_2016}. However, to the best of our knowledge there has not been a study to focus on the concept of luck and randomness applied to Risk, except to that of winning the game itself \cite{honary_total_2007}. 

This brief read of the literature suggests that luck is the relevant factors in a number of occasions, although there is a recognition that individual talent and skill may play a marginal role in some specific cases. As such, talent would be just a marginally weighting pendulum, however insufficient to counter-balance chance. 

We believe that the difficult in making a distinction of the causal relevance for either luck or talent derives from the complexity of quantitatively measuring them. Hence, a controlled simulation in which both: talent and luck are measured, along with a given context and a number of declared goals is a reasonable way to disaggregate this influencing force.

The contribution of this paper is to design and control a competitive, complex simulation with accountable doses of luck and, thus, be able to distinctly separate the influence of each component of success attainment. Further, we help conceptualize talent as a strategy, context as a simulation environment and perspective as a goal and apply those on a rule-based, reproducible, data-rich environment.

The literature suggests that the decomposition of each of the influences on success may be useful and applied to areas as diverse as finance, political forecasting or wage determination. 

Our results confirm previous literature towards the relevance of luck and randomness on success. However, we point further to the changing power of talent and context to achieve success. 

Besides this introduction, we present a description of the simulation, the rules, the strategies and the goals. Then we elaborate some hypothesis on the probabilities of the results, present them and discuss the results relative to the conceptualization made. We sum up with final remarks.

\section{\label{sec:sec2}Methods}

The procedures of the game follow a general version of the board game Risk, without distributing cards at the end of each turn. Basically, there are always six autonomous players that dispute 42 countries available in the world. Countries have a fixed set of neighbors and are organized within six continents. Although the simulation reasonably replicates the unfolding of a typical game with human players, the sole objective of the simulation is the capacity to significantly separate the characteristics of talent, context and perspective from luck.  

Before the first turn, the 42 countries are randomly assigned to each player so that each one has 7 countries before the game starts. Players also choose a random goal and strategy. Players then deploy their allotted amount of armies at once, and each, in turn, attacks. Whenever a country owner is attacked, he or she defends itself. After the first turn, army deployment happens before each players turn. Whenever there is a winner, the game is over.

\subsection{\label{sec:procedures}Procedures, rules, and a few details}
We implemented the simulation in \texttt{python}. You can find the code repository at \href{https://github.com/BAFurtado/HISMP}{GitHub.com/BAFurtado/Talent\_vs\_Luck}. We set countries as a \texttt{class} of their own listing their current owner, list of neighbors and continent affiliation. World is also set up as a class that controls the unfolding of the game. As such, it contains, players, countries, current turn, whether there is a winner, as well as the network structure of the countries. 

The other relevant \texttt{class} is that of players. It contains needed methods, such as \texttt{add} and \texttt{remove} countries, calculate army size, deploy them according to their strategies, keeping a record of owned countries and their armies. 

At each turn, players need to (i) define their priorities, (ii) allocate armies accordingly, (iii) decide where to attack and (iv) for how long. When they are done attacking, they need to (v) decide whether and how to rearrange possible remaining armies before letting the next player go. 

\paragraph{Battle.} A battle to take over a country is defined on luck -- rolling dice. Attackers can only attack when they have more than a single army in any given country. Attacks happen exclusively on neighboring countries. The size of the army exceeding one is the number of dice the attacker rolls, always with an upper bound of three. Defenders use the number of dice equivalent to the number of armies in their country, also limited to three. Applying these rules means that the number of dice rolling can be different for attackers and defenders. 

Dice on both sides -- attacking and defending -- are sorted and compared in pairs. Ties favor the defender. Losses are the result of each pair comparison. Thus, the highest value of the attack is compared to the highest value in the defense. Each loss, resulting from each pair comparison, reflects in one army either from attacker or defender withdrawn from the board. 

Note that armies are withdrawn only on pair comparisons. Thus, if the attacker has 4 armies and the defender has 3 armies, the attacker will play three against three. In this case, the sum of losses has to be three divided among the attacker and defender depending on who wins each sorted pair of comparison. However, when the attacker attacks with three and the defender defends with one and wins, for example, only one pair was compared and then just one army is subtracted from the attacker. 

Every pair on every attack is recorded for both players with $+1$ for the winner and $-1$ for the loser.

Once an attack decision has been made, battles go on as long as the attacker has a number of armies above one or the defender has at least one army standing. If successful the attacker, he or she occupies the newly conquered country with up to three armies. 

Each player at the beginning of each turn (except for turn 0) acquires a rule-based number of armies that can be allocated in any of the player's countries. Armies are equal to the number of owned countries (floor) divided by 2, or the scalar 3, whichever is larger.

Additionally, players who owns all countries for a given continent acquires armies to deploy within that continent, following the proportion: 7 armies for Asia, 5 armies for North America and Europe, 3 for Africa and 2 for South America and Australia.

\subsection{\label{strategies}Talent and Perspective}

\paragraph{Priorities.} The rationale that pushes forward each autonomous player reflects what human players usually do: seek continent completion, whatever their goal, so that the surplus of armies can propel them towards their objective. Typically, becoming a strong player is the only pathway towards achieving a goal. 

In the simulation, priority refers to a tuple of owned country--country to attack that are neighbors. Each player at each turn first calculates the proportion of each continent they own. Then, they will seek to conquer the continents starting at the one where they own the highest proportion. 

There are three alternative strategies randomly drawn to each player at each game start. They represent the inherent talent of which player. The strategies differ only on two moments. The first on (i) deciding how many armies to allocate on priority countries and the second on (ii) rearranging armies at the end of each turn. 

\paragraph{Strategies.} \textbf{Sensible}: is a strategy in which -- at allocating armies -- a maximum of three armies are allocated at each priority country. Further, armies are not rearranged, they simply remain at their location at the end of the turn when there are no more neighbors to attack.

\textbf{Minimalist}: is a strategy in which a single victory is aimed at on each turn. Once a new country is conquered, the player will stop. Also, when rearranging armies around countries, minimalist strategy will opt for overall protection maintaining a minimum of two armies at each country. 

\textbf{Blitz}: is a strategy in which the player will concentrate its armies on the first priority country. Also, when rearranging, blitz will gather all armies above one and distribute them on the priorities, enforcing protection on hubs. 

\paragraph{Goals.} The goals provide the individual perspective on deciding when there is a winner. At every end of turn, players check whether the goal has been achieved or not. There are four different goals, drawn randomly in the beginning of the game.

\textbf{Territory24} is a flexible goal in which possession of any 24 countries is sufficient to make a winner and -- in the context of the simulation -- determine success.

\textbf{Territory18} determines that the player needs to have possession of any 18 different countries with a further restriction of having at least 2 armies in each one of them.

\textbf{Continent} establishes the possession of all countries in four different combinations of continents which are drawn at the beginning of the game. They player may have to conquer all countries in Asia and South America or Asia and Africa or North America and Africa or North America, Asia and Australia. 

\textbf{Destroy} is to eliminate one of the other opponents so that the enemy's countries are zero. Although others may weaken the opponent, if the goal holder is not the one to take the last country of the enemy, its goal automatically changes to Territory24. 

\paragraph{Simulation, ties and reproducibility.} The user may run the simulation for a single game or thousands of times cloning the repository (link at section \ref{sec:procedures}). On a terminal, type \texttt{python game\_on.py} to run a single game. The simulation outputs: the evolution of each turn; players leaving the game -- due to having no countries left --; the name of the winner; its number of armies, strategy, goal and enemy; if any. Additionally, a \textit{gif} animation of the evolution of the graph of countries' possessions is saved. 

Most games will finish within the first tens of turns. We impose an \textit{ad hoc} limit of 200 turns when we stop the simulation and register a \textbf{tie}. In such cases, we collect information on the player with most countries at the time as the winner.

The user can run thousands of simulations \footnote{Default is 10,000 times, but 100,000 will take less than two hours to run.} typing \texttt{python simulation.py}. In this case, the output is a \textit{CSV} file containing the details of all runs.

The file records (i) the number of wins for each strategy-and-goal pair and whether there was a tie or not. Hence, for each triple combination the table shows: (ii) the number of countries the winner possessed, (iii) the average number of wins on each pair of dice rolling for the winner and (iv) for the average of the other players; (v) the number of pair comparisons for the winner and (vi) for the average of other players, along with additional supporting data for the second best player.

\section{\label{hypothesis}Hypothesis}
In this section we build some general hypothesis of expected results that will function as a null model in the results and analysis discussion that follow. 

Were the \textbf{strategies} considered to be independent and irrelevant to success, that is, if the game winner were dependent on luck alone; then we would expect an even distribution of winning strategies. Thus, each strategy would have an estimated probability of success of $1/3$. 

The probabilities for the goals are trickier. First, a player cannot have a goal of destroying itself. Thus, there is a $1/6$ chance that a player who gets 'Destroy' gets himself or herself as an enemy. In such cases, the goal is reassigned equally either to 'Territory24' or 'Territory18'. Thus, we deduce $1/24$ from 'Destroy' and add $1/48$ to both 'Territory' goals.

Second, the list of possible 'Continent' assignment contains only four combinations of continents. Thus, on the odd case that the $5^{th}$ or $6^{th}$ game goal pick is 'Continent', repeating at least the four previous picks as 'Continent', there are no more available combinations. When that happens, the goal is also randomly reassigned to either 'Territory18' or 'Territory24'. Therefore, we deduce $1/1024$ for each chance that there are no more continents to pick, and add $1/1024$ to each of the 'Territory' goals. 

Third, dynamically along the game, most of the players with the goal 'Destroy' may still become a 'Territory24' when they are not the ones to conquer the last country of their enemy. That is, when another player destroy the player's enemy, his or her goal is automatically changed. That further favors the goal 'Territory24'. For this case, we assumed that, on average, two out of the other four players will get to destroy your enemy before you, further deducing $1/40$ from the 'Destroy' goal, and adding it to 'Territory24'.

Starting from equal shares of $1/4$ goals, and adding and deducing the amounts above, we get a rough estimated probability that each goal amounts to the following percentage of the total runs: 'Territory24': 0.297, 'Territory18': 0.272, 'Continent': 0.248, 'Destroy': 0.183. 

Further, goals cannot be considered of the same degree of difficulty, nor independent. First, if we consider difficulty to be the number of countries to conquer, than 'Continent' can have lists of sizes: 15, 16, 18 and 25. That is, 'Continent' alone can vary among these numbers with an average of 18.5 countries. 

Moreover, there is first the issue of conquering a specific 'localized' country against 'any' country -- how to balance this difference in difficulty? Then, there is the point of accounting for providing one extra army on each country such as the goal of 'Territory18'. A third question would be how to quantify the fact that others may help you along the way, conquering your enemy's countries, however, simultaneously hindering you from winning when you are not the player to conquer the last standing country?

The issue of trying to differentiate goals' difficulties contributes to our concept of 'Context'. Success is given by circumstances and may be easier or harder to accomplish.

In this section we were able only to roughly estimate a probability of distribution of assigned goals and confirm their heterogeneity in difficulty. In the next section we will get the number of goals at the end of the game numerically for the 100,000 runs.

\section{\label{results}Results}

Results refer to a simulation of 100,000 games played. The statistics presented here come from running \texttt{simulation.py} with the option of generating new data to \texttt{False}. The reproducible data of the 100,000 runs is available at the GitHub repository.

\subsection{Talent}

Strategies are clearly relevant to success with results departing strongly from a hypothetical equal distribution of $1/3$. 

The 'Blitz' strategy dominates the results with 52,563 wins overall (see table \ref{tab:tab1}). That is 19 percentage points above the expected results. Strategy 'Sensible' comes next with 25,346 and 'Minimalist' is the least competitive with only 22,091 wins, which is nearly 11 percentage points below expected. Results suggest that performance is not marginally, but significantly different from average expected results.

Hence, for the remainder of the results, we implicitly consider 'Blitz' to be the most talented strategy, followed by 'Sensible' as a somewhat neutral strategy, and then 'Minimalist' as a less efficient strategy.

\begin{table}[!bh]
\caption{\label{tab:tab1} Number of wins by strategy out of 100,000 simulations and percentage points difference from hypothetical expected value of $1/3$.}
\begin{ruledtabular}
\begin{tabular}{@{}llr@{}}
\textbf{Strategy} & \textbf{Wins} & \textbf{Perc. pts diff.}\\
Blitz                      & 52,563  & 19 p.p. \\
Sensible                   & 25,346  & -8 p.p. \\
Minimalist                 & 22,091  & -11 p.p. \\
\end{tabular}
\end{ruledtabular}
\end{table}

\subsection{Context: talent + luck + perspective}

\begin{figure}[b]
\includegraphics[width=\linewidth]{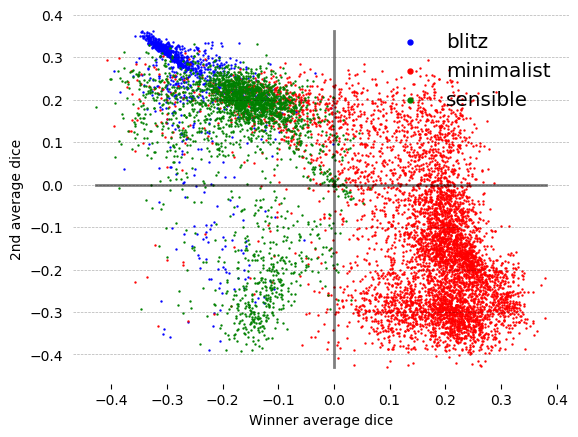}
\caption{\label{fig:fig1tietrue}Separation of luck, context and strategy for best and second-best players in games ending with a tie. Strategy 'Minimalist' has the most countries when the dice are mostly positive on average. The few times 'Blitz' and 'Sensible' are the winners, they do so even against luck. }
\end{figure}

Out of 100,000 runs only 9,973 did not end before the $200^{th}$ turn and were defined as 'Ties' (see table \ref{tab:tab2} and figure \ref{fig:fig1tietrue}). Nearly $2/3$ of the wins when there was a tie, 6,214, were achieved with the 'Minimalist' strategy for the player with most countries at the end.

\begin{table}[!bh]
\caption{\label{tab:tab2} Number of wins by strategy and whether there was a tie.}
\begin{ruledtabular}
\begin{tabular}{@{}llr@{}}
\textbf{Tie}   & \textbf{Strategy}   & \textbf{Wins}  \\
False & Blitz      & 51,556 \\
      & Minimalist & 15,877 \\
      & Sensible   & 22,594 \\
Total &            & 90,027 \\
True  & Blitz      & 1,007  \\
      & Minimalist & 6,214  \\
      & Sensible   & 2,752  \\
Total &            & 9,973 \\
\end{tabular}
\end{ruledtabular}
\end{table}

The numbers on table \ref{tab:tab2} suggest that there may be a strategy best suited for winning ('Blitz') and one for defending ('Minimalist'). However, when luck is added to the figure, it is clear that 'Minimalist' is good in keeping the game going only when luck is in its favor. Figure \ref{fig:fig1tietrue} shows the 9,973 simulations in which the game ended with a tie and the average dice for winners and the runner-ups. 'Blitz' and 'Sensible' in turn are strategies able to score wins in the long run even when their overall average is negative.

\begin{figure}[b]
\includegraphics[width=\linewidth]{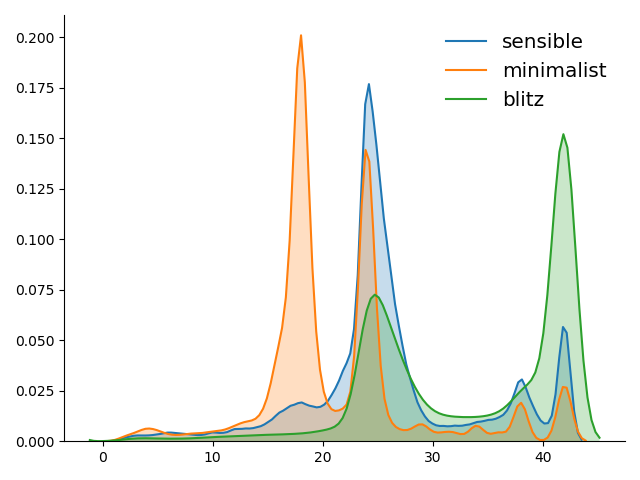}
\caption{Number of countries at end of simulation for each strategy}
\label{fig:fig2num_countries_strategy}
\end{figure}

Indeed, 'Blitz' seem to be much more aggressive than the other two strategies with a higher mean number of countries in the end of the game, compared to a much smaller number for strategies 'Minimalist', and 'Sensible' (see figure \ref{fig:fig2num_countries_strategy}). Moreover 'Minimalist' strategists seem to be more dependent to goal, scoring nearly half of its few wins at the mark of 18 conquered (and protected) countries. 

Conversely, 'Blitz' strategy seems incompatible with goal 'Territory18', winning only with dominance of the total number of countries in the game (see figure \ref{fig:fig2num_countries_strategy}). Only after having conquered everything, the player can build up its defense and have two or more armies in each one of 18 countries. Moreover, this aggressiveness is also shown on the average number of players still alive at the end of the game when the winner has strategy 'Blitz', compared to 'Minimalist' and 'Sensible'. 

\subsection{Perspective}

Relative to the results obtained for goals, 'Territory24' was the winner goal in 30,846 runs, which is very close to the estimated number of 0.2968\% of expected players holding such goal (Table \ref{tab:tab3}). 'Territory18', 'Destroy' and 'Continent' also perform close to their expected null results with respectively 26,716, 19,353 and 23,085 wins.

\begin{table}[!bh]
\caption{\label{tab:tab3} Actual results, percentage and hypothetical number of wins for each goal.}
\begin{ruledtabular}
\begin{tabular}{@{}lrrr@{}}
\textbf{Goals} & \textbf{Wins} & \textbf{Percentage} & \textbf{Hypothetical Expected}  \\
Continent           & 23,085  & 0.231	& 0.248 \\
Destroy             & 19,353  & 0.194	& 0.183 \\
Territory18         & 26,716  & 0.267	& 0.272 \\
Territory24         & 30,846  & 0.309	& 0.297 \\
\end{tabular}
\end{ruledtabular}
\end{table}

\begin{figure}[b]
\includegraphics[width=\linewidth]{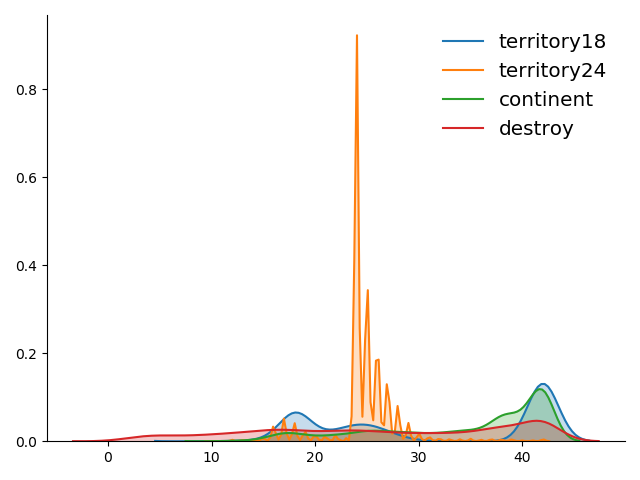}
\caption{\label{fig:num_countries_goal} Histogram of the number of countries at end of simulation for each goal}
\end{figure}

Another way to grasp the difficulty of the goals is to observe the number of countries at the end of the simulation for each one (see figure \ref{fig:num_countries_goal}). Note, for example, that 'Destroy' is a goal that is more evenly distribute. This indicates that the strength of the winner (number of its countries) is less important relatively to its goal. 'Territory18' and 'Continent' in turn seem to suggest that the goal is only achieved when shear dominance is conquered. 'Territory24' is the only goal that seem to lead to the abrupt conclusion of the game.

\subsection{Luck}

Despite the relevance of players' strategies, luck is determinant for the winner. The median average dice for winners over all runs is 0.071 (Table \ref{tab:tab4}). This compares with a median average of all the other players' of -0.058. The runner-up -- that player that comes in second -- gets a nearly neutral result at -0.013. In short, that means the winner is the luckiest one, on average. 

\begin{table}[!bh]
\caption{\label{tab:tab4} Median of Average Dice by Success.}
\begin{ruledtabular}
\begin{tabular}{@{}lrrr@{}}
\textbf{Average Dice} & \textbf{Median} & \textbf{No Tie} & \textbf{Tie}\\
Winner   & 0.071 & 0.070 & 0.084 \\
Runner-up & -0.013 & -0.013 & -0.014 \\
Others   & -0.058 & -0.055 & -0.102 \\
\end{tabular}
\end{ruledtabular}
\end{table}

\begin{figure}[b]
\begin{subfigure}[b]{1\linewidth}
\includegraphics[width=\linewidth]{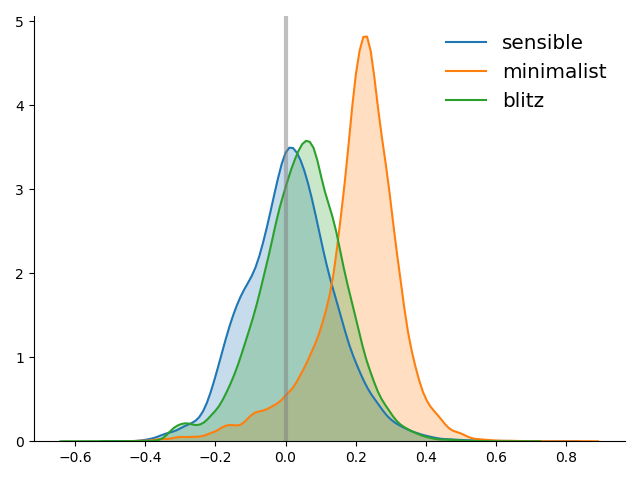}
\caption{Histogram of Winners' Average Dice by Strategy.}
\end{subfigure}
\begin{subfigure}[b]{1\linewidth}
\includegraphics[width=\linewidth]{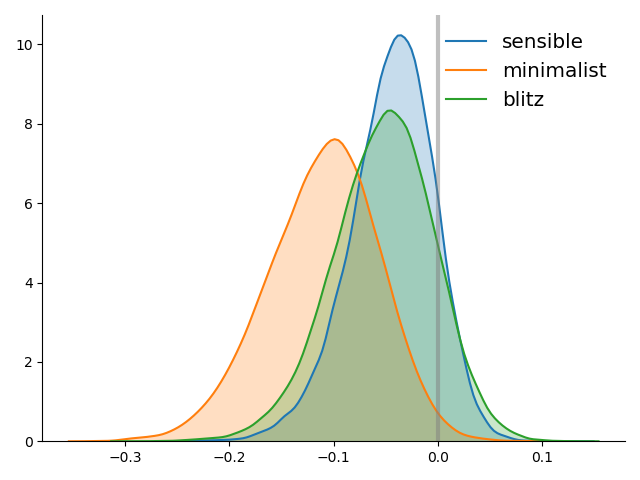}
\caption{Histogram of Other Players' Average Dice, given Winner's Strategy.}
\end{subfigure}
\caption{\label{fig:winner_avg}Distribution of Winners' and Other Players' average Dice relative to Winners' Strategy.}
\end{figure}

What differs, however, is the dependency on luck across different talent categories. In other words, although luck is fundamental, talent allows each player to be more or less dependent on it (see figure \ref{fig:winner_avg} and also figure \ref{fig:fig1tietrue}). For example, whereas the player with the 'Minimalist' strategy wins when its dice reaches a median value as high as 0.202, winning with 'Blitz' requires a value of only 0.048, which is just above a neutral value. 'Sensible' strategy, although winning much less often, does so with just a slightly positive luck of 0.015.

Alternatively, one may consider the median average luck of the players who lost for a given strategy. The inverted logic prevails. When a 'Minimalist' strategy wins, the other players' median average dice is very unlucky at -0.11. Such bad luck is a bit less pronounced for the other players when the winner plays 'Blitz', at -0.052 or 'Sensible', at -0.044. All in all, for a 'Minimalist' strategy win to happen a combination of winner's luck and opponents' bad luck also has to occur.

Talent seems to be relevant even to maintain a stalemate when luck is not favoring you. Table \ref{tab:tab5} shows that the only time winners have a negative median average happens when there is a 'Tie', which means they have been playing for up to 200 turns. At that point, winner is the player who owns more countries. 

Such negative value happens only for the two 'strongest' strategies: 'Blitz' and 'Sensible'. Surprisingly enough, the runner-up -- the contestant maintaining balance -- is on average much luckier (Table \ref{tab:tab5}) than the actual winner.

\begin{table}[b]
\caption{\label{tab:tab5} Luck of Winners, Runner-ups and Other Players by Strategy of the Winner and whether it was a Tie or not.}
\begin{ruledtabular}
\begin{tabular}{@{}llrrr@{}}
& & \textbf{Winners} & \textbf{Runner-ups} & \textbf{Other Players} \\
\textbf{Tie} & \textbf{Strategy} &  &  & \\
False  & Blitz          & 0.054     & -0.011 & -0.050  \\
             & Minimalist     & 0.236     & -0.114 & -0.098 \\
             & Sensible       & 0.029     & 0.039  & -0.040 \\
True  & Blitz & \textbf{-0.280}    & \textbf{0.289}  & -0.036 \\
 & Minimalist & \textbf{0.189}      & \textbf{-0.159} & -0.130 \\
 & Sensible & \textbf{-0.157}    & \textbf{0.185}  & -0.055
\end{tabular}
\end{ruledtabular}
\end{table}

\paragraph{Resilience and Opportunity.} 

Finally, it is noteworthy to point out that each players' talent may interfere with the number of times they engage in luck-dependent contexts. Table \ref{tab:tab6} shows that winners roll the dice about double the time of the other players. Further, more talented players are able to keep successful even when suffering bad luck for a very long period. For example, 'Blitz' players, when 'Tie' is true, have a very high number of dice-rolling (5,511 -- compared to an overall median of 242), which result in negative results (-0.281), and yet, they are able to finish the game with the highest number of countries. That also happens with the 'Sensible' strategy. Conversely, the 'Minimalist' player -- the least successful or talented of the strategies -- needs to maintain a positive luck streak in order to grasp success.

\begin{table}[b]
\caption{\label{tab:tab6} Number of Winners and Other Players' Dice Rolls and Dice Results by Strategy for the Occurrence or Absence of Ties. The data suggests that when 'Tie' is True for both strategies 'Blitz' and 'Sensible' success comes after a high number of Dice rolling and even with negative luck.}
\begin{ruledtabular}
\begin{tabular}{@{}llrrrrr@{}}
 & & \textbf{Wins}  & \textbf{W.Rolls} & \textbf{O.Rolls} & \textbf{W.Dice} & \textbf{O.Dice} \\
\textbf{Strategy}   & \textbf{Tie}   &  & & & & \\
Blitz      & False & 51,556 & 211     & 90    & 0.054  & -0.050 \\
Blitz      & True  & 1,007  & 5,511    & 1,620  & \textbf{-0.281} & -0.036 \\
Minimalist & False & 15,877 & 399     & 222   & 0.236  & -0.098 \\
Minimalist & True  & 6,214  & 3,961  & 1,258  & \textbf{0.189}  & -0.130 \\
Sensible   & False & 22,594 & 189     & 110   & 0.029  & -0.040 \\
Sensible   & True  & 2,752  & 3,461  & 1,576  & \textbf{-0.157} & -0.055
\end{tabular}
\end{ruledtabular}
\end{table}

\section{Discussion}
Given the conceptualization we gave at the introduction (section \ref{sec:sec1}) and the results presented, we make a case for the following disaggregation, given the context of a competitive sum-zero game with the rules and procedures described. 

\paragraph{Talent} -- taken as different mind-setting or different set of choices within the rule-based alternatives -- is relevant. Talent makes all the difference in the game. Which strategy the player chooses is likely to make a difference on the results. This difference, conditional to the rules of this simulation, may be as large as 19 percentage points, for the case of 'Blitz', for example, or at least minus 8 p.p. for the other two available choice. 

\begin{figure}[b]
\includegraphics[width=\linewidth]{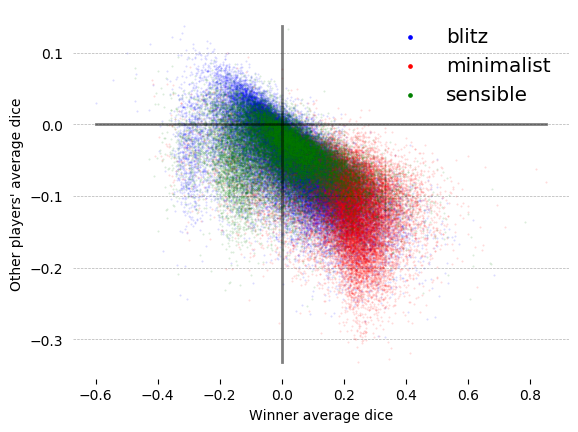}
\caption{\label{fig:fig3winners_others} Separation of luck, context and strategy. Clearly, different talent yields winners on distinct levels of luck, context and perspective.}
\end{figure}

\paragraph{Context} -- taken as the environment of which game, that is, other players' luck, goals and strategies -- is relevant. Figure \ref{fig:fig3winners_others} clearly shows that luck and strategy are correlated. A simple ANOVA between the strategies and winner's dice gets an F score of 15,896.529 with a p-value of 0.0. That means that a strategy may be the winning one, depending on how the context evolves, the right setting. Anecdotally, suppose you know to be running on a lucky streak, 'Minimalist' may be enough. A rainy day may only bring you success when playing 'Blitz', though. Wanting to play safe? Go for 'Sensible' even though with a larger chance of not getting your fair value.

\paragraph{Perspective} -- taken as the hardship and definition of attaining success -- is irrelevant. Despite the difficulty to elaborate hypothesis on the distinction among the available goals and how hard they may be to accomplish, yet, they seemed to matter very little. The expected number of wins was divided among the goals closely to how they had been constructed and distributed in the first place. Moreover, the explicit difference in number of countries to conquer -- between an average of 'Continent' of 18.5 to 24 from 'Territory24' seemed to be unimportant, relative to the power of the strategy. That was clearly demonstrated by the fact that a strong player, that is, with the talented strategy, led to wins independently of the number of countries needed for a given goal. \footnote{Were the strategies and goals totally random, we would expect any combination of strategy and goal to win $1/12$ or 0.083\% of the times.} 

\paragraph{Luck} is central! Despite the possibility of having disaggregated all the previous concepts with a reasonable differentiation, there is no doubt that luck is of the essence. Nearly 96\% of all wins have winners average dice as positive and above the value of the other players' dice values. The only exception are the 4,062 cases in which there is a tie, and the winner has much worse luck when compared to the runner-up or the other players.

In addition to the concepts designed to be investigated, two other ideas appear in the simulation as the analysis unfold. Investigating the number of rolls of each player, their strategy (talent) and their luck, we see that \textit{strategy determines Opportunity}. That is in the sense that a talented player may hold bad luck for longer periods and still have a shot at being successful. 'Minimalist' players, on the contrary, would maintain competitiveness only when being extremely lucky. 

A somewhat loose narrative might just be that the strategy allows for the player to wait out as much as needed, until luck strikes. That implies that a given \textit{strategy is more correlated to Resilience} than others (see the dominance of nearly $2/3$ of 'Minimalist' when considering ties).  

\section{Final remarks}
We design a controlled, ruled-oriented, open and reproducible simulation based on the Risk board game in order to examine different contributions from luck, context and perspective on success. We found the results to be informative even considered the limitations of this specific design. 

We argue that luck, talent and context are relevant to success, whereas perspective -- or the description of success -- is not. In fact, the results suggest that luck is central in the simulation. Even though luck is key, the talent the player has also seems to be determinant. More so within different contexts. Further, we learned that talent may even offer more opportunities for luck to strike. In doing so, talent has also proved to be more resilient and able to remain competitive for much longer periods. Finally, the highest the talent, less dependent on luck the player is.

In sum, this paper contributes to current literature as a way of quantifying an explicit disaggregation of the proposed concepts. We also confirm previous findings and extend the exercise to include context and perspective, with also some hinting into resilience and opportunity. 

Additionally, If one considers a financial perspective and associate the 'Blitz' strategy to an agent who is risk-taking, the 'Sensible' strategy as a risk-neutral agent and 'Minimalist' as a risk-averse agent; then we also replicate stylized-facts of finance. That is, you need to take more risks to win in the long-run. 

For future work, we are considering two extensions. One is to add 'cards' to the game. Simply, they add an extra layer of luck to the game, making it more complex, closer to actual played games and one more chance to associate card playing with strategy. We may also consider using the known results of different strategies as an \textit{a priori} construction of talent and make a distribution of strategy more likely to the ones observed in reality. That is, fewer players would have higher strategy, with the majority of them having an average-quality strategy.

\begin{acknowledgments}
We wish to acknowledge the support of grant number [306954/2016-8] from the National Council of Research (CNPq), Brazil.
\end{acknowledgments}

\bibliographystyle{apsrev4-2}
\bibliography{TvsLuck}
\end{document}